\documentstyle[12pt,epsf]{article}

\newcommand\be{\begin{equation}}
\newcommand\ee{\end{equation}}
\newcommand\bea{\begin{eqnarray}}
\newcommand\eea{\end{eqnarray}}
\newcommand\beas{\begin{eqnarray*}}
\newcommand\eeas{\end{eqnarray*}}

\newcommand\lae{\stackrel{<}{\sim}}
\newcommand\gae{\stackrel{>}{\sim}}

\newcommand\iic{\ \ ,}
\newcommand\iip{\ \ .}
\newcommand\sha{\hat s}
\newcommand\tha{\hat t}
\newcommand\uha{\hat u}
\newcommand\snum{\sha - M_c^2}
\newcommand\unum{\uha - M_c^2}
\newcommand\tnum{\tha - M_c^2}
\newcommand\sden{\vert \sha_c \vert^2}
\newcommand\tden{\vert \tha_c \vert^2}
\newcommand\uden{\vert \uha_c \vert^2}
\newcommand\shac{\sha - M_c^2 + i\Gamma(s) M_c}

\newcommand\stnum{(\snum)(\tnum) + \Gamma^2(s) M_c^2}
\newcommand\tunum{(\tnum)(\unum) + \Gamma^2(s) M_c^2}
\newcommand\tr{{\rm Tr}}
\newcommand\gev{{\rm GeV}}

\begin{document}

\begin{titlepage}
\def\thepage {}     

\title{Coloron phenomenology}
\author{Elizabeth H. Simmons\thanks{e-mail address:
simmons@bu.edu} \\
Department of Physics, Boston University \\
590 Commonwealth Ave., Boston  MA  02215}

\date{\today}
\maketitle

\bigskip
\begin{picture}(0,0)(0,0)
\put(295,250){BUHEP-96-24}
\put(295,235){hep-ph/9608269}
\end{picture}
\vspace{24pt}

\begin{abstract}
A flavor-universal extension of the strong interactions was recently
proposed in response to the apparent excess of high-$E_T$ jets in the
inclusive jet spectrum measured at the Tevatron.  This paper studies
the color octet of massive gauge bosons (`colorons') that is present
in the low-energy spectrum of the model's Higgs phase.  Constraints
from searches for new particles decaying to dijets and from
measurements of the weak-interaction $\rho$ parameter imply that the
colorons must have masses greater than 870-1000 GeV.  The implications
of recent Tevatron data and the prospective input from future
experiments are also discussed.
  
\pagestyle{empty}
\end{abstract}

\end{titlepage}


A flavor-universal coloron model \cite{newint} was recently proposed 
to explain the apparent excess of high-$E_T$ jets in the 
inclusive jet spectrum measured by CDF \cite{CDFexc}.  This model is a
flavor-universal variant of the coloron model of Hill and Parke
 \cite{topglu} which can accommodate the jet excess without
contradicting other experimental data.  It involves a minimal extension
of the standard description of the strong interactions, including the
addition of one gauge interaction and a scalar multiplet, but no new
fermions.  Furthermore, the flavor-universal coloron model of the
strong interactions can be grafted onto the standard one-Higgs-doublet
model of electroweak physics, yielding a simple, complete, and
renormalizable theory.   

The model serves as a useful baseline with which to compare both the
data and other candidate explanations of the jet excess.  The latter
include such diverse physics as modified parton distribution functions
\cite{pdfs}, higher-order contributions from gluon resummation
\cite{hogr},  phenomenological models of quark
substructure \cite{CDFexc}, new strongly-coupled $Z^\prime$ gauge bosons
\cite{othr2}, quark resonances \cite{othr3}, non-standard triple-gauge
vertices \cite{othr4}, and light gluinos  \cite{othr5}.

This letter explores the phenomenology of the Higgs phase of the model,
in which an octet of strongly-interacting massive gauge bosons
(colorons) is present in the low-energy spectrum.   Previous work on
this model has considered effects on the inclusive jet spectrum
\cite{newint} and the dijet spectrum and angular distributions
\cite{rmh} in the approximation where coloron exchange is treated as
a four-fermion contact interaction.  The four-fermion approximation is 
appropriate for heavy colorons and conveniently allows the model's
effects to be represented in terms of a single parameter: the
coefficient of the contact interaction.  The present work goes beyond
the four-fermion approximation by including propagating colorons of
finite width.  This allows us to investigate the phenomenology of
colorons too light to be well-described by contact interactions at
Tevatron energies.  It also enables us to explore the separate
dimensions of the coloron parameter space (mass and mixing angle)
independently.

We begin by reviewing the main features of the model, in both the Higgs
and confining phases.  Then, focusing on the Higgs phase, we discuss
the effect that the presence of colorons strongly coupled
to quarks will have on partonic scattering cross-sections.  In section
\ref{sec:lim-now}, we show what limits may already  be placed on the
mass and mixing angle of the colorons.  Next, we discuss how further
experiments can probe the model at higher coloron masses. Our
conclusions appear in section \ref{sec:concl}.

\section{The model}
\label{sec:model}
\setcounter{equation}{0}

In the flavor-universal coloron model  \cite{newint}, the strong gauge
group is extended to $SU(3)_1 \times SU(3)_2$.  The gauge couplings
are, respectively, $\xi_1$ and $\xi_2$ with $\xi_1 \ll \xi_2$.  Each
quark transforms as a (1,3) under this extended strong gauge group.

The model also includes a scalar boson $\Phi$ transforming as
a $(3,\bar 3)$ under the two $SU(3)$ groups.  The most
general\footnote{As noted in  \cite{newint} this model can be grafted
onto the standard one-doublet Higgs model.  In this case, the most
general renormalizable potential for $\Phi$ and the Higgs doublet
$\phi$ also includes the term $\lambda_3 \phi^\dagger\phi
\tr(\Phi^\dagger\Phi)$.  For a range of $\lambda$'s and parameters in
the Higgs potential, the vacuum will break the two $SU(3)$ groups to
QCD and also break the electroweak symmetry as required.} potential for
$\Phi$ is
\be
U(\Phi) = \lambda_1 \tr\left(\Phi\Phi^\dagger - f^2 {\rm I}\right)^2 +
\lambda_2 \tr\left(\Phi\Phi^\dagger - \frac 13 {\rm I} \left(\tr
\Phi\Phi^\dagger\right)\right)^2
\ee
where the overall constant has been adjusted so that the minimum of $U$
is zero.  For $\lambda_1,\,\lambda_2,\,f^2 > 0$ the scalar develops a
vacuum expectation value $\langle\Phi\rangle = {\rm diag}(f,f,f)$
which breaks the two strong groups to their diagonal subgroup.  We
identify this unbroken subgroup with QCD.

The original gauge bosons mix to form an octet of massless gluons and
an octet of massive colorons.  The gluons interact with quarks
through a conventional QCD coupling with strength $g_3$.  The colorons
$(C^{\mu a})$ interact with quarks through a new QCD-like coupling
\be
{\cal L} = - g_3  \cot\theta J^a_\mu C^{\mu a} \iic
\ee
where  $J^a_\mu$ is the color current
\be
\sum_f {\bar q}_f \gamma_\mu \frac{\lambda^a}{2}q_f \iip
\ee 
and $\cot\theta = \xi_2/\xi_1\, $.  Note that we expect $\cot\theta
>1$.  In terms of the QCD coupling, the gauge boson mixing angle and
the scalar vacuum expectation value, the mass of the colorons is
\be
M_c = \left( \frac{g_3}{\sin\theta \cos\theta} \right) f \iip
\ee
The colorons decay to all sufficiently light quarks; assuming
there are $n$ flavors lighter than $M_c/2$, the decay width is
\be
\Gamma_c \approx \frac n6 \alpha_s \cot^2\theta\, M_c
\label{eqwid}
\ee
where $\alpha_s \equiv g_3^2/4\pi$. We take the top quark mass to be
175 GeV so that n = 5 for $M_c \lae 350$ GeV and n = 6 otherwise.

Thus far, we have described the Higgs phase of the model, in which 
$SU(3)_1\times SU(3)_2$ breaks at an energy scale where neither gauge
coupling is strong.  If the $SU(3)_2$ gauge coupling becomes strong
while the gauge symmetry is intact, the model enters a confining phase
instead.  As discussed in  \cite{newint}, if the strong coupling does
not break the chiral symmetries of the quarks, then it will bind the
quarks and the scalar $\Phi$ into $SU(3)_2$-neutral states. 
These composite fermionic states would correspond to the
`ordinary quarks' seen at low energies.  The remainder of this paper
will concentrate on the phenomenology of the Higgs phase of the model;
for more on the confining phase's phenomenology see ref.
 \cite{newint}. 

\section{Partonic cross-sections involving colorons}
\label{sec:partonic}
\setcounter{equation}{0}

Production and exchange of the massive colorons will affect hadronic
cross-sections.  To quantify this observation, we need to 
calculate the effect of the colorons on each partonic scattering
sub-process.  Two-body scattering involving external gluons remains the
same as in QCD; the relevant cross-sections may be found in
 \cite{QCDcr}.  Processes involving external quarks are modified by
coloron exchange and the cross-sections are given below.

For each two-body parton scattering cross-section, we write
\be
\frac{d\sigma}{d \tha} (a b \to c d) = \frac{\pi \alpha_s^2}{\sha^2}
\Sigma(a b \to c d) 
\ee
to define the $\Sigma(a b \to c d)$, which conventionally include
initial state color averaging factors and are written in terms of the
partonic Mandelstam invariants $\sha, \tha,$ and $\uha$.  For
scattering of light quarks, whose masses may be neglected, we have
\be
\Sigma(q q^\prime \to q q^\prime) = \frac 49 (\sha^2 + \uha^2)
\left[ \frac{1}{\tha^2} + \frac{2 \cot^2\theta (\tnum)}{\tha\ \tden} +
\frac{\cot^4\theta}{\tden} \right]
\ee
\be
\Sigma(q \bar q \to q^\prime {\bar q}^\prime) = 
\frac 49 (\tha^2 + \uha^2)
\left[ \frac{1}{\sha^2} + \frac{2 \cot^2\theta (\snum)}{\sha\ \sden} +
\frac{\cot^4\theta}{\sden} \right]
\label{eqsig}
\ee
\bea
\Sigma(q q \to q q) &=& \frac 49 (\sha^2 + \uha^2)
\left[ \frac{1}{\tha^2} + \frac{2 \cot^2\theta (\tnum)}{\tha\ \tden} +
\frac{\cot^4\theta}{\tden} \right]  \\
&+& \frac 49 (\sha^2 + \tha^2)
\left[ \frac{1}{\uha^2} + \frac{2 \cot^2\theta (\unum)}{\uha\ \uden} +
\frac{\cot^4\theta}{\uden} \right] \nonumber \\
&-& \frac{8}{27} \sha^2 \left[ \frac{1}{\tha\uha} + \frac{\cot^2\theta
(\tnum)}{\uha\ \tden} + \frac{\cot^2\theta (\unum)}{\tha\ \uden}
\right. \nonumber\\ 
& &\ \ \ \ + \left.\frac{\cot^4\theta \left[ \tunum\right]}{\tden
\uden}\right] \nonumber
\eea
\bea
\Sigma(q \bar q \to q \bar q) &=& \frac 49 (\sha^2 + \uha^2)
\left[ \frac{1}{\tha^2} + \frac{2 \cot^2\theta (\tnum)}{\tha\ \tden} +
\frac{\cot^4\theta}{\tden} \right] \\
&+& \frac 49 (\tha^2 + \uha^2)
\left[ \frac{1}{\sha^2} + \frac{2 \cot^2\theta (\snum)}{\sha\ \sden} +
\frac{\cot^4\theta}{\sden} \right] \nonumber \\
&-& \frac{8}{27} \uha^2 \left[ \frac{1}{\sha\tha} +
\frac{\cot^2\theta (\snum)}{\tha\ \sden} + \frac{\cot^2\theta
(\tnum)}{\sha\ \tden}
\right.
\nonumber \\ 
& &\ \ \ \ + \left. \frac{\cot^4\theta \left[\stnum\right]}{\sden\
\tden} \right] \nonumber
\eea
where $q^\prime$ denotes a quark of a flavor other than $q$, and where
$\sha_c \equiv \shac$ and similar definitions hold for $\tha_c$ and
$\uha_c$.  In the above expressions, $\Gamma(s)$ is the $s$-dependent
width of the coloron
\bea
\Gamma(s) &\approx& \frac{n}{6}\cot^2\theta \sqrt{\sha}\ \ \ \ \ \
\sha < M_c^2  \nonumber \\
&=& \Gamma_c \ \ \ \ \ \ \sha > M_c^2
\eea

The top quark is heavy enough to warrant
separate treatment.  The proton's top quark content is negligible and
we need not consider processes involving initial-state top quarks. 
Furthermore, top quarks are produced in quark-quark scattering with a
mass-dependent cross-section  \cite{combr}; when the contributions of
colorons are included, we find
\bea
\Sigma(q\bar q \to t \bar t) &=& \frac 49 (\tha^2 + \uha^2 + 
4 m_t^2\sha - 2 m_t^4) \ \times \\
& & \ \ \ \ \ \ \left[ \frac{1}{\sha^2} + \frac{2 \cot^2\theta
(\snum)}{\sha\ \sden} + \frac{\cot^4\theta}{\sden} \right]\iip \nonumber
\eea
The cross-section for $gg \to t\bar t$ is the same as in QCD and may
be found in ref.  \cite{combr}.

\section{Existing limits on colorons}
\label{sec:lim-now}
\setcounter{equation}{0}

We now describe the limits that may already be placed
on the mass and mixing angle of the colorons.  Scattering data from
hadron colliders and measurements of the weak-interaction $\rho$
parameter provide experimental input.  The condition that the model
be in the Higgs phase imposes a further bound on the mixing angle.

\subsection{dijets}

\begin{figure}[hb]
\vspace{-3.5cm}
\epsfxsize 10cm \centerline{\epsffile{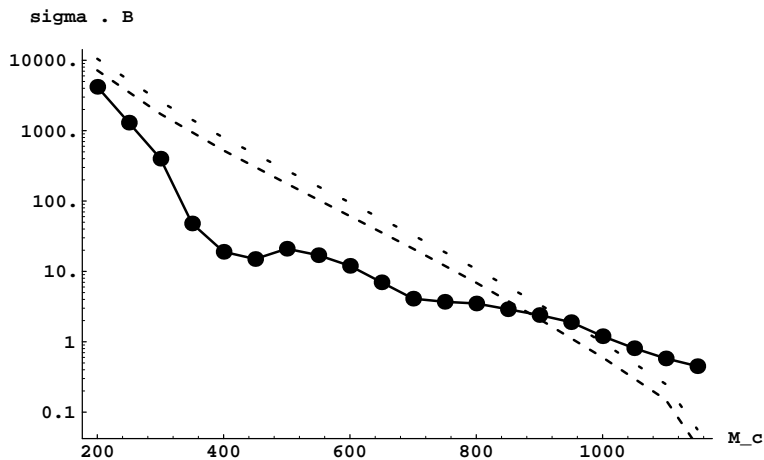}}
\vspace{-3.5cm}
\caption[dijetlim]{Experimental  \cite{CDFdij} upper limit on the
cross-section times branching ratio for new particles decaying to dijets
(points) is compared to theoretical predictions for colorons with
$\cot\theta = 1$ (lower dashed curve) and $\cot\theta = 1.5$ (upper
dashed curve).  Both jets are required to have pseudorapidity
$\vert\eta\vert < 2.0$ and the dijet satisfies $\vert\cos\theta^*\vert
< 2/3$. The experimental curve is based on 19pb$^{-1}$ of data.}
\label{dijetlim}
\end{figure}

A sufficiently light coloron would be visible in direct production at
the Tevatron.  Indeed, the CDF Collaboration has searched for new
particles decaying to dijets and reported  \cite{CDFdij} an
upper limit on the incoherent production of such states. 
Accordingly, we calculated $\sigma \cdot B$ for colorons with various 
values of $M_c$ and $\cot\theta$; the results for $\cot\theta = 1, 1.5$
are shown in figure \ref{dijetlim} together with the CDF limit.  In
calculating $\sigma\cdot B$, we followed the example of CDF in using
CTEQ structure functions and in requiring $\vert\eta\vert < 2$ and
$\vert\cos\theta^*\vert < 2/3$.  For $\cot^2\theta < 2$, the coloron's
half-width falls within the dijet mass resolution of 10\%; for larger
$\cot\theta$ we counted only the portion of the signal that falls
within a bin centered on the coloron mass and with a width equal to the
resolution.  

Values of $M_c$ and $\cot\theta$ which yield a theoretical prediction
that exceeds the CDF upper limit are deemed to be excluded at 95\%
c.l.  \cite{CDFdij}. We find that for $\cot\theta = 1$, the range $200
{\rm GeV} < M_c < 870 {\rm GeV}$ is excluded; at $\cot\theta = 1.5$, the
upper limit of the excluded region rises to roughly 950 GeV; at
$\cot\theta = 2$, it rises to roughly 1 TeV.  As the coloron width
grows like $\cot^2\theta$, going to higher values of
$\cot\theta$ does not appreciably increase the upper limit of the
excluded range of masses beyond 1 TeV.  

We can extend the excluded range of coloron masses to values below
those probed by CDF by noting two things.  First, $\sigma$ increases
as $\cot\theta$ does, so that exclusion of $\cot\theta = 1$ for a given
$M_c$ implies exclusion of all higher values of $\cot\theta$ at that
$M_c$.  Second, $\sigma \cdot B$ is the same for a coloron with
$\cot\theta = 1$ as for an axigluon\footnote{Axigluons \cite{axig} are
the strongly coupled massive gauge bosons that remain in the spectrum
after spontaneous breaking of an $SU(3)_L \times SU(3)_R$ gauge group
to the vector subgroup identified with QCD.} of identical
mass\footnote{Thus our limit on $M_c$ with
$\cot\theta = 1$ agrees numerically with the limit CDF \cite{CDFdij} 
gives for the axigluon mass.}, as one may verify by comparing our
equations (\ref{eqwid}) and (\ref{eqsig}) with equations (1.1) and
(2.2) in ref.
 \cite{axigphenom}.  Axigluons with masses between 150 and 310 GeV have
already been excluded by UA1's analysis  \cite{ua1} of incoherent
axigluon production; by extension, colorons in this mass range with
$\cot\theta \geq 1$ are also excluded.  

\medskip

The  combined excluded ranges of $M_c$ are
\bea
150 \gev < &M_c& < 870 \gev \ \ \ \ \ \ \cot\theta = 1\nonumber \\
150 \gev < &M_c& < 950 \gev \ \ \ \ \ \ \cot\theta = 1.5 \\
150 \gev < &M_c& < 1000 \gev \ \ \ \ \ \cot\theta \gae 2\iip\nonumber
\eea
They are summarized by the shaded region of figure \ref{allcurlim}

\subsection{b-tagged dijets}

Because the colorons couple to all flavors of quarks, they should also
affect the sample of b-tagged dijets observed at Tevatron
experiments.  The CDF Collaboration has reported  \cite{bbdij} limits on
certain new particles decaying to b-tagged dijets, based on the 1992-3
run.   Their limit on narrow topgluons indicates the kind of limit that
the data would provide for colorons.  

Briefly, topgluons \cite{topglu} belong to an $SU(3)_1 \times SU(3)_2$
model in which light quarks transform as (3,1)'s while $t$ and
$b$ are (1,3)'s.  When the $SU(3)_1\times SU(3)_2$ breaks to its 
diagonal QCD subgroup, an octet of gauge bosons becomes massive.  These
topgluons, $B^{\mu a}$, couple to quarks as  \cite{topglu}
\be
{\cal L} = -\left[g_3 \cot\theta_t \left( \bar t \gamma_\mu
\frac{\lambda_A}2 t + \bar b \gamma_\mu \frac{\lambda_A}2 b \right) -
g_3 \tan\theta_t J^a_\mu \right] B^{\mu a}
\ee
where $q_i$ are light quarks, $\theta_t$ is the mixing angle between
topgluons and ordinary gluons, and $\cot\theta_t \gg 1$. Because of the
factor of $\cot\theta_t$ in the first term and $\tan\theta_t$ in the
second, the process $\ q\bar q \to topgluon \to b\bar b\ $ is
independent of
$\theta_t$ (except where it enters through the topgluon width).  This
contrasts with
$b\bar b$ production through coloron exchange, which grows as
$\cot^4\theta$.  Furthermore, heavy topgluons decay essentially only to
$t\bar t$ or $b\bar b$, while heavy colorons decay to all six quark
flavors.   In other words, (e.g. for topgluons and colorons heavier
than $2m_t$) 
\bea
\sigma_{topgluon} &\approx& {\sigma_{coloron}}/{\cot^4\theta}
\nonumber \\
B_{topgluon \to b\bar b} &=& 1/2 \\
B_{coloron \to b\bar b} &=&  1/6 \nonumber
\eea
which implies that $(\sigma \cdot B)_{topgluon}$ and $(\sigma\cdot
B)_{coloron}$ are equal for equal-mass bosons when $\cot^4\theta
\approx 1.7$.   Using equation (\ref{eqwid}), this corresponds to $
\Gamma_c/M_c \approx 0.13$.

CDF's range of excluded masses for narrow ($\Gamma/M = 0.11$)
topgluons of $200\, {\rm GeV} < M_{topgluon} < 500\, {\rm GeV}$ 
therefore provides a rough idea of the limit for colorons with
$\cot^2\theta \approx 1.7$.  It appears that the limit on colorons from
b-tagged dijets will be weaker than that from the full dijet sample.  
This contrasts with the case of topgluons, which can be more strongly
constrained by the b-tagged dijet sample because they decay almost
exclusively to third-generation quarks.

\subsection{the $\rho$ parameter}

An additional limit on the coloron mixing angle may be derived from
constraints on the size of the weak-interaction $\rho$-parameter. 
Coloron exchange across virtual quark loops contributes to $\Delta\rho$
through the isospin-splitting provided by the difference between the
masses of the top and bottom quarks.  Limits on this type of
correction  \cite{cdt} imply that
 \cite{newint}
\be
\frac{M_c}{\cot\theta} \gae 450 {\rm GeV} \iip
\label{rhoweq}
\ee
This excludes the hatched region of the $\cot^2\theta$ -- $M_c$
plane shown in figure \ref{allcurlim}.  Note that this excludes an
area of small $M_c$ that the dijet limits did not probe, as well as an
area at larger $M_c$ and large $\cot\theta$.

\subsection{critical value of $\cot\theta$}

Finally, we mention a theoretical limit on the coloron parameter
space.  While the model assumes $\cot\theta > 1$, the value of
$\cot\theta$ cannot be arbitrarily large if the model is to be in the
Higgs phase at low energies.   Writing the relationship between the
couplings $\xi_1$ and $\xi_2$  of the original gauge groups and the QCD
coupling $g_3$ as
\be
\xi_1 \cos\theta = g_3 = \xi_2 \sin\theta
\ee
confirms that $\xi_2$ is large when $\cot\theta$ is large.  If
$\cot\theta$ is large enough, then $\xi_2$ exceeds its critical value,
and the low-energy theory is in the confining phase rather than the
Higgs phase.  

\begin{figure}[htb]
\vspace{-3.5cm}
\epsfxsize 10cm \centerline{\epsffile{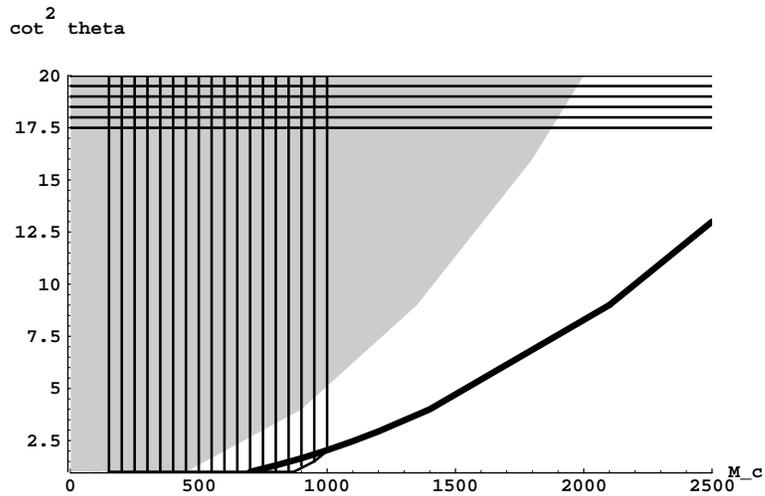}}
\vspace{-3.5cm}
\caption[allcurlim]{Current limits on the coloron parameter space: mass ($M_c$)
vs. mixing parameter ($\cot^2\theta$).  The shaded region is excluded
by the weak-interaction $\rho$ parameter \cite{newint} as in equation
\ref{rhoweq}.  The vertically-hatched polygon is excluded by searches
for new particles decaying to dijets \cite{CDFdij,ua1}. The
horizontally-hatched region at large $\cot^2\theta$ lies outside the
Higgs phase of the model.  The dark line is the curve $M_c /
\cot\theta = 700$ GeV for reference.} 
\vspace{-.5cm}
\label{allcurlim}
\end{figure}

Writing the low-energy interaction among quarks that results
from coloron exchange as a four-fermion interaction
\be
{\cal L}_{4f} = - \frac{g_3^2 \cot^2\theta}{M_c^2} J^a_\mu J^{a \mu}
\label{fourff}
\ee
we use the NJL approximation to estimate the critical value of
$\cot^2\theta$ as\footnote{In more conventional notation (see e.g.
\cite{topglu}), one would write the coefficient of the four-fermion
operator as $-(4\pi\kappa/M^2)$ and find $\kappa_{crit} = 2 \pi/ 3$.}
\be
(\cot^2\theta)_ {crit} = \frac{2\pi}{3 \alpha_s}  \approx 17.5
\ee
This puts an upper limit on the $\cot^2\theta$ axis of the coloron's 
parameter space, as indicated in figure \ref{allcurlim}.

\section{Prospective limits}
\label{sec:lim-when}
\setcounter{equation}{0}

Data from run IA and IB at the Tevatron, from future Tevatron runs,
and eventually from the LHC have the potential to shed further light on
the flavor-universal coloron model.  We suggest below what sort of
analysis may prove useful.

\subsection{jet spectra at the Tevatron}

\begin{figure}[htb]
\vspace{-3.5cm}
\epsfxsize 10cm \centerline{\epsffile{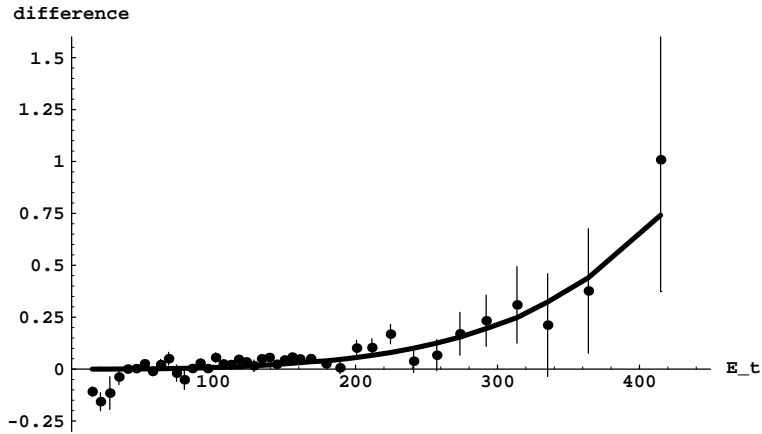}}
\vspace{-3.5cm}
\caption[differ]{Difference plot ((data - theory)/theory) for the
inclusive jet cross-section ${1\over{\Delta\eta}} \int
(d^2\sigma/d\eta\, dE_T) d\eta $ as a function of transverse jet
energy $E_t$, where the pseudorapidity $\eta$ of the jet falls in the
range $0.1 \leq \vert\eta\vert \leq 0.7$. Dots with (statistical)
error bars are the recently published CDF data \protect\cite{CDFexc}.
The solid curve shows the LO prediction of QCD plus the contact
interaction approximation to coloron exchange of equation
(\protect\ref{fourff}) with $M_C/\cot\theta = 700$ GeV.  Following
CDF, we employed the MRSD0' structure functions \protect\cite{mrs_pak}
and normalized the curves to the data in the region where the
effect of the contact interactions is small (here this region is $45 <
E_T < 95$ GeV).}
\label{diffc}
\end{figure}

Both the inclusive jet  spectrum ($d\sigma/ d E_T$) and the dijet
invariant mass spectrum ($d\sigma / d M_{jj}$) measured in CDF's run IA
and IB data  \cite{CDFexc} appear to show excesses at high energy end of
the spectrum.  The dijet limits we derived earlier imply that the
coloron is heavy enough that it would not be directly produced in
the existing Tevatron data.  Therefore, it is useful to start studying
the data in terms of the four-fermion approximation (\ref{fourff}) to
coloron exchange. Comparison with the run IA CDF inclusive jet spectrum
already  \cite{newint} indicates that $M_c / \cot\theta = 700$ GeV is not
obviously ruled out, as figure \ref{diffc} illustrates. Figure
\ref{allcurlim} indicates where the curve $M_c / \cot\theta = 700$ GeV
falls relative to the limits on the parameter space discussed earlier. 
Detailed analysis including both systematic and statistical errors
should be able to determine a lower bound on $M_c / \cot\theta$. 

For colorons weighing a little more than a TeV -- those that are
just above the current dijet mass bound -- it is more appropriate to
use the cross-sections for full coloron exchange (see section
\ref{sec:partonic}) when making comparisons with the data.   Such
colorons are light enough that their inclusion yields a
cross-section of noticeably different shape than the four-fermion
approximation would give (see figure 4).  Once the full
coloron-exchange cross-sections are employed, the mass and mixing angle
of the coloron may be varied independently.  In particular,  one may
study the effects of light colorons with small values of
$\cot^2\theta$.  This can expand the range of accessible parameter
space beyond what one would have reached by using the four-fermion
approximation.

\begin{figure}[htb]
\vspace{-3.5cm}
\epsfxsize 10cm \centerline{\epsffile{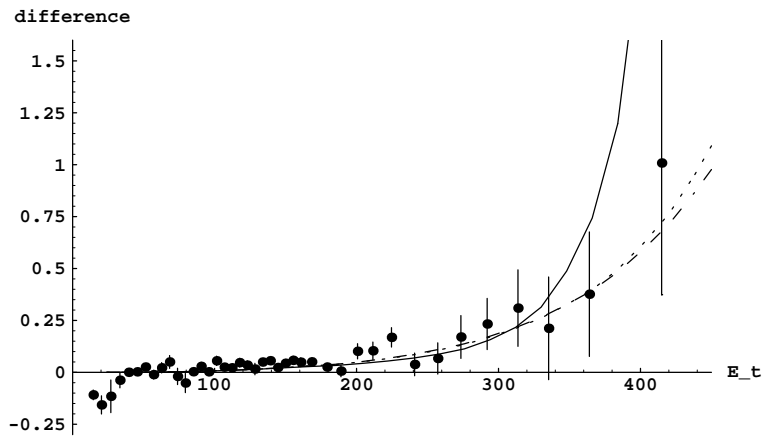}}
\vspace{-3.5cm}
\caption[shape]{Difference plot for $d\sigma / d E_T$ (see 
figure 3) showing the effects of colorons of different masses when the
ratio $M_c / \cot\theta$ is fixed at 700 GeV.  Here full one-coloron
exchange is included, rather than the contact interaction
approximation.  The solid curve is for a light coloron: $M_c = 1050$
GeV, $\cot\theta = 1.5$.  The dotted and dashed curves correspond to much
heavier colorons ($M_c = 1750$ GeV and 2000 GeV) with correspondingly
larger values of $\cot\theta$ (2.5 and 3.0).  The cross-section for
the heavier colorons is well-approximated by the contact interaction 
approximation at Tevatron energies; the cross-section for the lighter coloron
is not.}
\label{shape}
\end{figure}

\subsection{dijet angular distributions at the Tevatron}

Another means of determining what kind of new strong
interaction is being detected is measurement of the dijet angular
distribution.  Some new interactions would produce dijet angular
distributions like that of QCD; others predict distributions of
different shape.  In terms of the angular variable
$\chi$ %
\be
\chi = \frac{1 + \vert\cos\theta^*\vert}{1 - \vert\cos\theta^*\vert}
\ee
QCD-like jet distributions appear rather flat while those which are
more isotropic in $\cos\theta^*$ peak at low $\chi$ (recall that
$\theta^*$ is the angle between the proton and jet directions).   The
ratio $R_\chi$ 
\be
R_\chi \equiv \frac{N_{events}, 1.0 < \chi < 2.5}{N_{events}, 2.5 < \chi
< 5.0}
\ee
then captures the shape of the distribution for a given sample of
events, e.g. at a particular dijet invariant mass.  

The CDF Collaboration has made a preliminary analysis of the dijet angular
information in terms of $R_\chi$ at several values of dijet invariant
mass \cite{rmh}.  The preliminary data appears to be consistent either with
QCD or with QCD plus a color-octet four-fermion interaction like
(\ref{fourff}) for  $M_c/\cot\theta = 700$ GeV.  Our calculation of
$R_\chi$ including a propagating coloron gives results consistent with
these.  It appears that the measured  angular distribution can allow the
presence of a coloron and can help put a lower bound on
$M_c/\cot\theta$. 

\subsection{heavy colorons at the lhc}

Measurement of the inclusive jet and dijet spectra at the LHC will
probe higher values of the coloron mass.  The existence of a critical
value of the mixing angle discussed earlier has an interesting
implication for LHC experiments.  Suppose analysis of Tevatron
inclusive jet and dijet data finds that the best fit comes from
including a four-fermion interaction of the form (\ref{fourff}) with
$M_c/\cot\theta = X$ GeV.  This four-fermion interaction can
self-consistently be described as the low-energy manifestation of
exchange of a massive coloron only if $M_c
\le X \cdot \cot\theta_{crit}$.   Thus the Higgs-phase coloron model
would predict that experiments at the LHC should find a coloron
resonance at or below that mass.  For example, figure 3 shows how a
four-fermion interaction with $X = 700$ GeV compares with the CDF
inclusive jet data; that would correspond to a coloron weighing no
more than about 3 TeV.

\section{Conclusions}
\label{sec:concl}
\setcounter{equation}{0}

The flavor-universal coloron model can accommodate an excess
at the high-$E_t$ end of the inclusive jet spectrum at Tevatron
energies without contradicting other data.  Previous measurements of the
weak-interaction $\rho$ parameter and searches for new particles
decaying to dijets imply that the coloron must have a mass of at least
870 GeV.  Ongoing and future experiments at hadron colliders have the
power to test the model further.  In particular, if Tevatron
experiments identify a preferred value of the ratio $M_c/\cot\theta$,
then the upper limit on $\cot\theta$ in the model's Higgs phase would
yield an upper limit on the coloron mass to guide searches at the LHC.

\bigskip
\centerline{\bf Acknowledgments}
\vspace{12pt}

We thank R.S. Chivukula and R.M. Harris for useful conversations and
comments on the manuscript.
E.H.S. thanks the Aspen Center for Physics and the Fermilab Summer
Visitors Program for hospitality during the completion of this work.
E.H.S. also acknowledges the support of the NSF Faculty Early Career
Development (CAREER) program and the DOE Outstanding Junior Investigator
program. {\em This work was supported in part by the National Science
Foundation under grant PHY-95-1249 and by the Department of Energy
under grant DE-FG02-91ER40676.}


\end{document}